\documentclass[floatfix,prl,twocolumn,showpacs,preprintnumbers,amsmath,amssymb]{revtex4}
\usepackage{graphicx}% Include figure files
\usepackage{dcolumn}% Align table columns on decimal point
\usepackage{bm}% bold math

\begin{document}

%\preprint{Institute For Nonlinear Science, 2002}

\title{Estimating good discrete partitions from observed data:\\
  symbolic false nearest neighbors}% Force line breaks with \\

\author{Matthew B. Kennel}
\email{mkennel@ucsd.edu}
\author{Michael Buhl}
 \email{mbuhl@ucsd.edu}
\affiliation{%
Institute For Nonlinear Science\\
University of California, San Diego\\
La Jolla, CA 92093-0402
}%

\date{April 2, 2002}

\begin{abstract}
  
  A symbolic analysis of observed time series data requires
  making a discrete partition of a continuous state space containing
  observations of the dynamics.  A particular kind of partition,
  called ``generating'', preserves all dynamical information of a
  deterministic map in the symbolic representation, but such
  partitions are not obvious beyond one dimension, and existing
  methods to find them require significant knowledge of the dynamical
  evolution operator or the spectrum of unstable periodic orbits.  We
  introduce a statistic and algorithm to refine empirical partitions
  for symbolic state reconstruction.  This method optimizes an
  essential property of a generating partition: avoiding topological
  degeneracies.  It requires only the observed time series and is
  sensible even in the presence of noise when no truly generating
  partition is possible.  Because of its resemblance to a geometrical
  statistic frequently used for reconstructing valid time-delay
  embeddings, we call the algorithm ``symbolic false nearest
  neighbors''.
  
\end{abstract}

\newcommand{\x}{{\bf x}}
\newcommand{\y}{{\bf y}}
\newcommand{\z}{{\bf z}}
\renewcommand{\P}{{\cal P}}
\newcommand{\N}{{\cal N}}
\newcommand{\Jsfnn}{J_{\text{sfnn}}}
\newcommand{\Ksfnn}{K_{\text{sfnn}}}
\newcommand{\Jupo}{J_{\text{upo}}}
\newcommand{\dht}{ {\delta h}_T }


\maketitle

Why might one want to represent observed time series of dynamical
systems as sequences of low-precision discrete symbols?  In this
representation, there are often interesting techniques---often (but
not exclusively) derived from information theory and its associated
technology---which may illuminate data in novel
ways~\cite{symbolictimeseriesanalysis}.  The initial step for all
these methods requires making a {\em partition}: a coloring of the
state space~\cite{reconstruction}, $\x \in R^d$, into non-overlapping
regions and associated symbols so that any $\x$ is assigned
a unique symbol $s$ in a discrete alphabet. % of size $A$. 
The symbol may be represented as an integer in the set ${0, 1, \ldots A-1}$.
A partition $\P$ defines a discretization of the observed
sequence $\x_i, i=1\ldots N$ into a symbolic sequence, $s_i, i=1\ldots
N$.

What partitions are ``good''?  Which discretizations retain the full
structure of the original dynamics in the $\x$ space in the sequence
of symbols?  Unfortunately, the situation is unlike the remarkable
time-delay embedding method for continuous dynamics: simple partitions
are {\em not} generically satisfactory.  The mathematics of symbolic
dynamics specifies what we want: a ``generating partition'' (GP),
where symbolic orbits uniquely identify one continuous space orbit,
and thus the symbolic dynamics is fully equivalent to the continuous
space dynamics.

Unfortunately there is no satisfactory mathematical theory about how
to find a GP as a general procedure (except for one dimensional
dynamics ($d=1$), where partitioning at the critical points works).
Are the ad-hoc partitions often used still satisfactory?
Unfortunately they are often not so.  Bollt {\em et al}~\cite{bollt}
examined the degradation in the symbolic dynamics which results from
the frequently used ``histogram partition'', as opposed to a GP.  A
less optimal partition will induce improper projections or
degeneracies, where a given symbolic segment may correspond to more
than one topologically distinct state space orbit.  This resulted in
finding the wrong topological entropy.  Chaotic communication with
symbolic targeting works most satisfactorily knowing a GP, because
then the transmitted symbolic message may be directly mapped into a
desired orbit in the attractor.  (see, e.g.  ~\cite{lai2dsaddle})

Since a partition is a critical first step for any symbolic data
analysis, a poor partition yield poor results, a method to approximate
good partitions from observed data {\em alone} is urgently needed.
With apparently no existing satisfactory solutions, this is the
problem we attack.  Davidchack {\em et al}~\cite{davidchack} recently
presented a partitioning method which works by successively coloring
unstable periodic orbits (UPOs) to ensure unique codings (all UPOs
have unique codes under a GP).  The necessary high-order UPOs are very
difficult to obtain from observed data alone, unfortunately.

The Kolmogorov-Sinai entropy rate $h_{KS}$ of the dynamics can be
found from the supremum, over all increasingly fine partitions, of
Shannon's entropy rate evaluated on the information source implied by
the discretization.  More strikingly, a GP also achieves this supremum
with a finite, and, one hopes, small alphabet.  This suggests a
naive strategy whereby one maximizes a statistical {\em estimator}
of the entropy rate evaluated on the sequence induced by candidate
partitions~\cite{ebeling}.  This apparently attractive idea is flawed
as demonstrated by the following counterexample.  Consider a partition
of the state space with a fine box-size $\epsilon$ where each region
is {\em randomly} assigned a symbol from the alphabet.  For
sufficiently small $\epsilon$, the symbol sequence of any finite time
series will appear indistinguishable from a memoryless and
structureless information source with the maximum possible entropy
rate $h = \log_2 A$, since each observed datum could have
encountered a different partition element with a new random symbol..
With the highest entropy possible this partition would be selected
over competitors but is clearly useless for data analysis as the
resulting symbolic stream says nothing about the original time series
or its particular dynamics.  Even if one believes this pathology to be
irrelevant to coarser encodings, there are practical problems with the
maximum-estimated-entropy idea.  First, estimation of $h$ is not
trivial to do well; second, when there is observational noise
(inevitable with data acquisition equipment) larger alphabets will
appear to give significantly higher entropies even if they are not
actually much better at encoding the dynamics.  As the true entropy
rate of the system is not already known (and often a key quantity one
wants to estimate {\em given} a good partition), there is no absolute
statistical target which confirms whether the proposed partition is at
all close or far from the ideal.  In practice, selecting partitions
with entropy does not seem to work well in general.

We assert our practical criterion for a good partition: {\em
  short sequences of consecutive symbols ought to localize
  the corresponding continuous state space point as well as possible.}
A good coding ought to maintain the benefits of a low-precision
symbolic representation with minimum distortion of the original state
space dynamics.  Our central idea is to form a particular geometrical
embedding of the symbolic sequence under the candidate partition and
evaluate, and minimize, a statistic which quantifies the apparent
errors in localizing state space points.

We embed the symbol sequence into the unit square~\cite{henonsymbols}:
\begin{equation}
\y_i =
\left( \sum_{k=1}^{k_{\text{max}}} s_{i-(k-1)} /A^k,
  \sum_{k=1}^{k_\text{max}} s_{i+k} / A^k  \label{eq:symbologram}
\right). 
\end{equation}
($k_{\text{max}}$ is chosen such that $A^{-k_{\text{max}}}$ is as
small as the computational precision.)  For a binary alphabet $(A=2)$,
the first coordinate of $\y_i$ is the binary fraction whose digits
start at $s_i$ and go backwards in time, the second is with the
sequence going forward from $s_{i+1}$.  Intuitively, the distribution
on $\y$ is like a $\P$-dependent symbolic version of the invariant measure.

%One potentially undesirable effect is the arithmetic ordering in the
%na\"ive symbolic plane.  For example the string ${\tt 01\bar{1}}$ will
%be mapped close to ${\tt 10\bar{0}}$ despite differing in all symbols.
%A Gray code ordering will alleviate this issue.  Each forward and
%backward string of symbols is considered a Gray code word and we
%substitute its arithmetic conversion.  For any symbol string $g_k, k=1
%\ldots L$, the digits of its arithmetic counterpart, $a_k[g_k]$, are
%$a_k = g_k$ when $(\Sigma_{k-1} \mod 2) = 0$ and $a_k = A-1-g_k$
%otherwise, with $\Sigma_k = \sum_{l=1}^k g_k$, $\Sigma_0
%= 0$. 
%With the required finite-range
%cutoff
%, the mapping $\phi$ is evaluated on the observed
%trajectory:
%\begin{equation}
%    \y_i = \left( \sum_{k=1}^{k_{\text{max}}} \frac{ a_k[s_{i-(k-1)}] }{ A^k}, 
%      \sum_{k=1}^{k_{\text{max}}} \frac{a_k[s_{i+k}]}{ A^k} \right). \label{eq:symbolgray}
%\end{equation}
%We do not presume that this coding always makes the $\y$ space
%dynamics very simple, as was the case for the particular map studied
%in~\cite{henonsymbols} and applied to communication
%in~\cite{lai2dsaddle}.  We need only define a sensible distance
%between itineraries.

Given $\x_i$ and a partition $\P$, the symbolic
embedding~(\ref{eq:symbologram}) yields a parallel series $\y_i$,
defining points on some map $\y = \phi_\P(\x)$.  We want $\phi_\P$ to
be {\em injective}, i.e.  $\phi_\P(\x) = \phi_\P(\x')$ implies
$\x=\x'$.  With finite data, we desire that if $||\phi_\P(\x) -
\phi_\P(\x')||$ is small, so is $||\x - \x'||$.  By construction,
sufficiently near points in $\x$ have close symbolic sequences in their
most significant digits.  In a good partition, additionally, nearby
points in $\y$ remain close when mapped back into the $\x$-space.  By
contrast, bad partitions induce topological degeneracies where similar
symbolic words map back to globally distant regions of state space.
As shown in~\cite{bollt}, this phenomenon confounds proper analysis of
the observed symbolic dynamics.

We need to quantify how well any candidate partition achieves our
ideal.  We find the {\em nearest neighbor}, in Euclidean distance, to
each point $\y_i$. Conventional \mbox{$k$-d} tree
algorithms~\cite{kdtree} efficiently provide the index of the nearest
neighbor to any point in a data set: $\N[i] = \text{arg} \min_{k \ne
  i} ||\y_k - \y_i ||$.  Knowing symbolic neighbors, we find distances
of those same points back in $\x$-space, $D_i = ||\x_{\N[i]} -\x_i||$.
We normalize the set of $D_i$ by a monotonic transformation: given any
$D$, find its rank $R \in [0,1]$ in the cumulative distribution of
{\em random} two-point distances $||\x_\alpha - \x_\beta||$.  Large
$R$ means that localizing well in symbol space did not localize well
in the original state space.

Better partitions give a smaller proportion of {\em symbolic false
  nearest neighbors}, that fraction of $R_i$ which are greater than
some threshold $\eta$, denoted $\Jsfnn$.  This resembles the false
neighbors statistic for time-delay embeddings~\cite{falseneighbors}:
both count large-deviation ``mistakes'' in a related space which
result from topological mis-embedding in the tested space.
Appropriate values for $\eta$ which defining a ``large'' deviation are
$\eta \approx 0.01 - 0.3$, depending on the noise in $\x_i$.  An
alternative to $\Jsfnn$ is $\Ksfnn$, defined as the arithmetic average of the
largest $\gamma$ percentile of the set of $R_i$.  Using $\Jsfnn$, $\eta$
may need tuning depending on the noise scale and dynamical system, but
the effect of changing $\gamma$ is lower.  On the downside, $\Ksfnn$
does not necessarily converge to near zero for the optimal partition.
We typically find good results with $\gamma \approx 0.01-0.05$.

\begin{figure}
\includegraphics[width=\columnwidth]{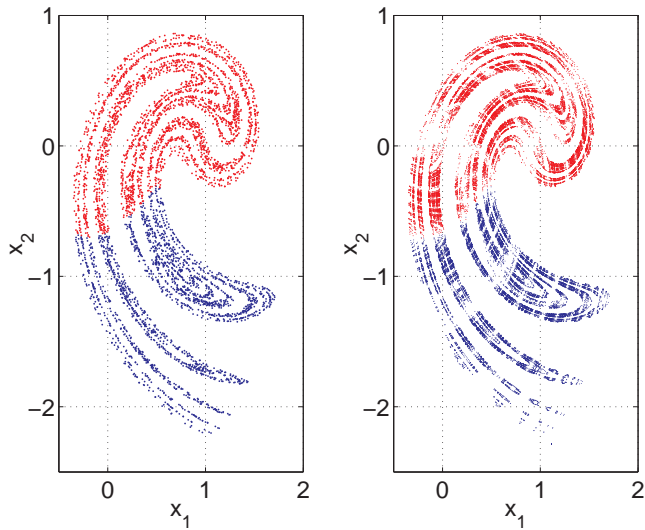} %figures/ikeda_partition.eps}
\caption{
  Left: partition estimated by SFNN optimization on 2000 data points
  from the Ikeda map.  Right: partition calculated with foreknowledge
  of UPOs, numerically extracted from the equation of motion.  The
  partition we estimate from observed data alone is quite close to a
  presumably correct one, calculated from the method
  of~\cite{davidchack}.  The measure on the two figures is not the
  same: the left figure is a sample of the natural measure, whereas
  the right shows UPOs up to period 16.  They avoid regions of
  homoclinic tangencies, contributing to the blank spaces.
  \label{fig1}}
\end{figure}

For concrete numerical calculations, we need to parameterize
partitions with a relatively small number of free parameters.
Inspired by~\cite{davidchack}, we define partitions with respect to a
set of radial-basis ``influence'' functions of the form $f_k(\x) =
\alpha_k / || \x - \z_k||^2$, the set of $\alpha$ and $\z$ being the
free variables.  For any particular $\x$, one $f_l(\x)$ will
generically result in the largest value versus other $f_k(\x), k \ne
l$, and then $\x$ is assigned to that symbol which was pre-assigned to
influence function $f_l$.  The $\z$ parameters are initialized to
random examples from the $\x_i$ and $\alpha$ to independent random
variates $[0,1)$, and $n_f$ functions assigned to each of the $A$
symbols.  In~\cite{davidchack} the $\z_k$ were fixed on the UPOs and
their symbols varied; here, the centers and coefficients vary but
their symbols are fixed.
%The candidate partition yields a mapping from $\x_i$ to a candidate
%$^s_i$. 
We minimize $\Jsfnn$ or $\Ksfnn$ over the $A n_f (d+1)$ free
parameters using ``differential evolution''~\cite{diffevol}, a genetic
algorithm suitable for continuous parameter spaces.  

Figure~\ref{fig1} shows the final ${\cal P}$ on 2000 data points from
the Ikeda map\cite{ikedamap}.  It shows the best result (lowest
$\Jsfnn$) out of six restarts changing only the random seed governing
the initial conditions; the results were not much worse on the other
runs, however.  The result is very close to the partition knowing
the dynamics. 

\begin{figure}
\includegraphics[height=2in]{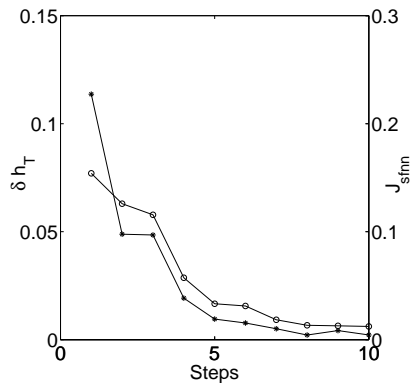} %figures/entropy.eps}
\caption{\label{fig2} For each new best partition:
  Minimization target $\Jsfnn$ (circles, and right scale), estimated
  deficiency in topological entropy $\dht$ (asterisks and left scale)
  Minimizing $\Jsfnn$ generally minimizes $\dht$ and thus maximizes
  topological entropy of the symbolic language.}
\end{figure}
\begin{figure}
\includegraphics[height=2.7in]{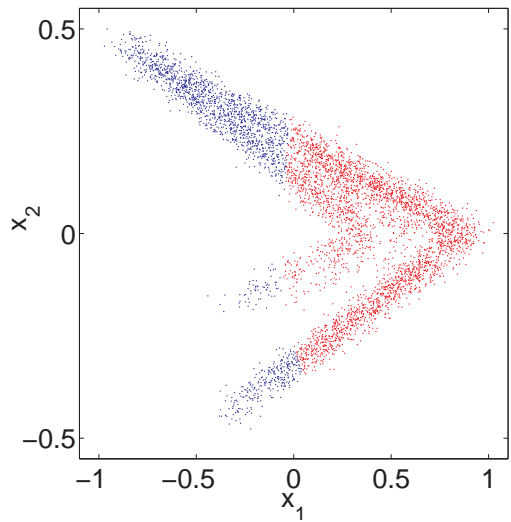} %figures/lozi_partition.eps}
\caption{\label{fig3} Minimizing $\Ksfnn$ with $\gamma=0.01$: estimated
  partition for a time series of 5000 data points from the Lozi map
  with 10\% additive by amplitude Gaussian noise.  Either the $x_1$ or
  $x_2$ axes are GPs for the noiseless map. Here
  despite the noise the algorithm finds a partition close to what
  would be a GP.}
\end{figure}
\begin{figure}
\includegraphics[height=3in]{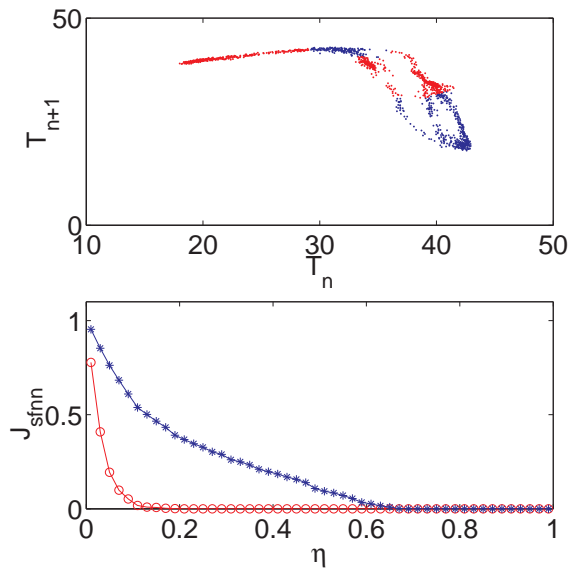} %figures/bubbles_partition.eps}
\caption{\label{fig4} Top: estimated binary partition for time-delay 
  embedding of inter-bubble time intervals (arbitrary units),
  minimizing $\Ksfnn$.  Bottom: $\Jsfnn(\eta) $ vs $\eta$ for the
  optimized partition (circles) and for a naive equiprobable histogram
  partition with the same alphabet (asterisks).  For the optimized
  partition there are very few large distance errors, e.g. $\Jsfnn$
  observed above $\eta = 0.1$. }
\end{figure}
\begin{figure}
\includegraphics[height=3in]{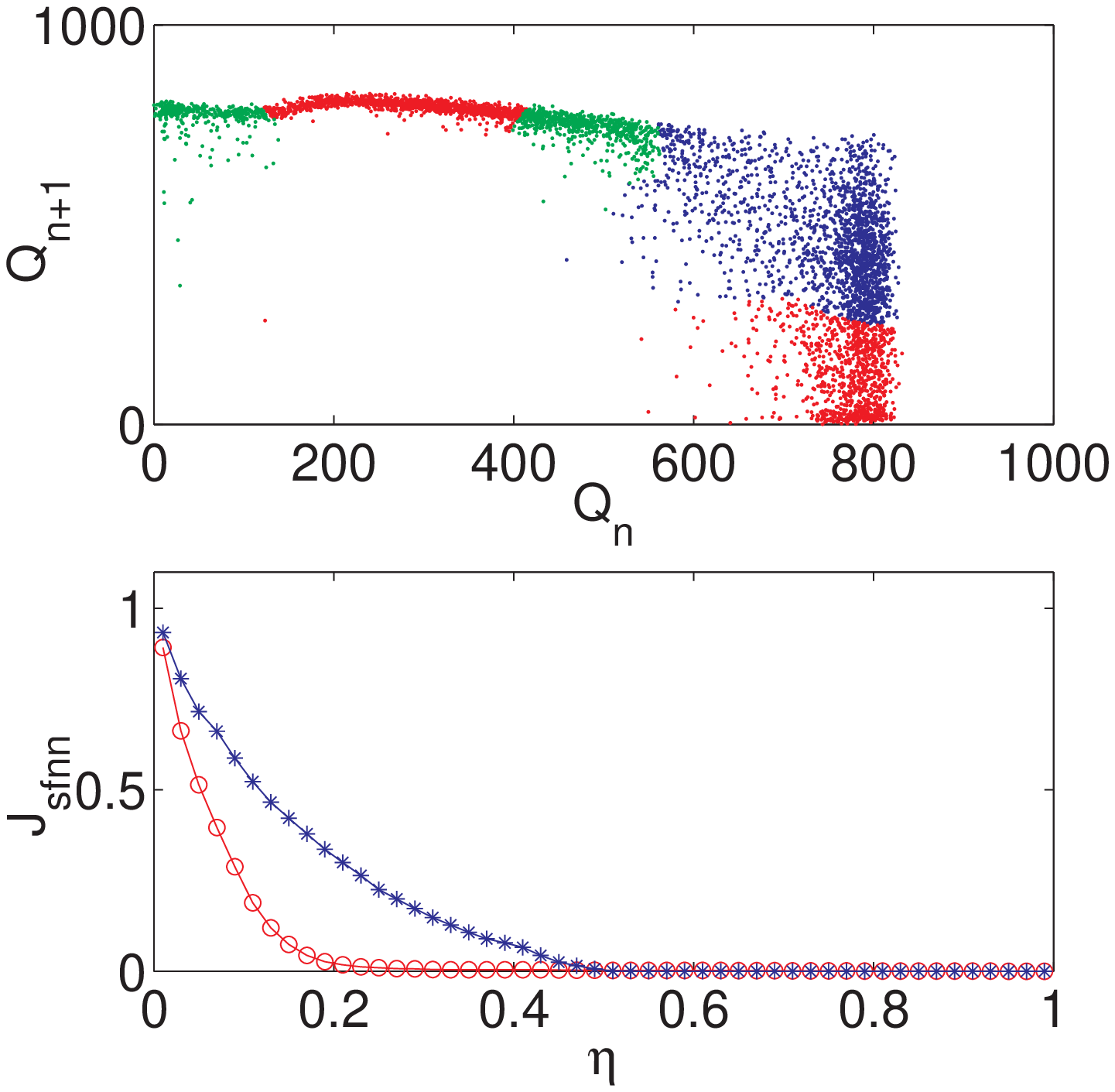} %ures/engine_partition.eps}
\caption{\label{fig5} Same as Fig.~\ref{fig4} but with combustion engine
  heat release time series (energy, arbitrary units), and $A = 3$.
  The noise level is higher thus there remain more moderately sized
  distances, even with a larger alphabet which usually results in
  better localization.}
\end{figure}

In a stationary information source, the number of distinct length-$p$
codewords will scale, for asymptotic $p$, as $N_p \propto e^{h_T p}$
where $h_T$ is the {\em topological entropy}, a dynamical invariant.
We validate ${\cal P}$ with an estimate of the deficiency between
$h_T$ implied by ${\cal P}$ and the correct $h_T$: $ \dht =
{p_\text{max}}^{-1} \sum_{p=1}^{p_\text{max}} {p}^{-1} \log \left(
  N_p/ \tilde{N}_p \right)$.  $N_p$ is the number of distinct
period-$p$ UPOs (which were computed knowing the equations of motion),
$\tilde{N}_p$ the number of such UPOs with unique $p$-symbol codes in
some ${\cal P}$.  A GP gives $\dht = 0$, and $\dht \rightarrow 0$ for
better (less UPO-degenerate) partitions.  Figure~\ref{fig2} shows
$\dht$ on each new best partition found during the optimization.  The
optimization target, $\Jsfnn$, decreases strictly monotonically by
construction; though $\dht$ does not decrease quite monotonically, the
trend toward very small values is clear.  This gives evidence that
minimizing $\Jsfnn$ also refines approximations to GPs.

Figures~\ref{fig3}--\ref{fig5} demonstrate applications of the
algorithm.  Fig.~\ref{fig3} shows the effect of noise on a system
where the GP is analytically known.  The Lozi map (see analysis
in~\cite{henonsymbols}) is similar to the Henon map but replaces the
quadratic nonlinearity with a piecewise linear one.  We find a
partition which is close to the noise-free GP even when the data have
been contaminated by significant amounts of additive noise.  Though
complete localization to a single point is not possible here,
minimizing large divergences is still a desirable criterion.
%(Dynamical noise would not likely produce ${\cal P}$ as close to the
%noise-free GP.)  
Figures~\ref{fig4} to \ref{fig5} show estimated
partitions on experimental data sets where no analytical form of the
equations (much less partitions) are known.  On account of noise,
dynamical or observational, a certain amount of divergence $D$ for
symbolic nearest neighbors is inevitable.  Still, minimizing large
deviations is a reasonable goal even for noisy data.  There are very
few rank distances with $R \ge 0.2$ or $0.3$ compared to a basic
histogram partition.

%The issue of proper symbolic reconstruction has been recognized but
%there are no good algorithms extant in the literature.  We believe the
%method of symbolic false nearest neighbors fulfills a pressing need in
%the symbolic analysis of observed nonlinear time series data: how to
%choose a partition that best maintains the dynamical structure of the
%system. One must keep in mind that when reconstructing the state space
%from data the embedding dimension in the \emph{continuous} space must be
%properly chosen ahead of time by alternative means
%(e.g.~\cite{falseneighbors}).  The present algorithm will attempt to
%find symbolic partitions that best localize to the continuous state
%space as given, but if that state space is improperly reconstructed,
%the resulting partition may still be far from generating even if it
%has few large symbolic nearest neighbor distances. 

It must be kept in mind that GPs are not necessarily unique for any
attractor.  Distinctly different partitions may be found, all of which
are reasonably satisfactory.  (At a minimum any iterate of a GP is
also a GP).  There are many coexisting solutions (roughly like finding
low-energy states of a spin glass), which is why the optimization
problem is hard, requiring a global search method.  We conjecture this
is one reason why understanding the structure of generating partitions
has been so difficult for mathematicians.

We also point out that it is possible to partition one space $\x$, but
quantify distances $D_i = ||\z_{\N[i]} -\z_i||$ in some different
space as long as there is some relation between each $\x_i$ and
$\z_i$.  For example, one may be interested in a simple symbolic
control scheme, say where $\z_i = \x_{i+T}$ (find the best partition
of observables now that best predicts some future), or perhaps when
two different variables are measured simultaneously and one wants to
cross-predict.  In the first case, in principle, a GP should be
optimal for predicting the future as well as the present but the
inevitable issues of finite data and noise may make the best empirical
partition different for the two cases.  In the second case, generating
partitions are irrelevant entirely.


\begin{thebibliography}{99}
\bibitem{ikedamap} Dynamical equations and parameters are as in
  \cite{davidchack}.
\end{thebibliography}
\end{document}